\documentclass[aps,pre,reprint,showpacs]{revtex4-2}

\usepackage{graphicx}
\usepackage{epstopdf}
\graphicspath{{figures/}}
\usepackage{amssymb}
\usepackage{amsmath}
\usepackage{color}
\usepackage{siunitx}
\usepackage{todonotes}

\begin{document}
\title{Lane formation in gravitationally driven colloid mixtures consisting of up to three different particle sizes}
\author{Kay Hofmann}
\altaffiliation{Physics Department, University of Konstanz, 78457 Konstanz, Germany; current address: Physics Department, University of Mainz, 55122 Mainz}
\author{Marc Isele}
\altaffiliation{Physics Department, University of Konstanz, 78457 Konstanz, Germany}
\author{Artur Erbe}
\altaffiliation{Helmholtz-Zentrum Dresden-Rossendorf, 01328 Dresden}
\author{Paul Leiderer}
\altaffiliation{Physics Department, University of Konstanz, 78457 Konstanz, Germany}
\author{Peter Nielaba}
\altaffiliation{Physics Department, University of Konstanz, 78457 Konstanz, Germany}
\email{peter.nielaba@uni-konstanz.de}

\begin{abstract}
Brownian dynamics simulations are utilized to study segregation phenomena far from thermodynamic equilibrium. In the present study, we expand upon the analysis of binary colloid mixtures and additionally introduce a third particle species to further our understanding of colloidal systems. Gravitationally driven, spherical colloids immersed in an implicit solvent are confined in two-dimensional linear microchannels. The interaction between the colloids is modeled by the Weeks-Chandler-Andersen potential, and the confinement of the colloids is realized by hard walls based on the solution of the Smoluchowski equation in half space. In binary and ternary colloidal systems, a difference in the driving force is achieved by differing colloid sizes, but fixed mass density. We observe for both the binary and ternary systems that a driving force difference induces a nonequilibrium phase transition to lanes. For ternary systems, we study the tendency of lane formation in dependence of the diameter of the medium-sized colloids. Here, we find a sweetspot for lane formation in ternary systems. Furthermore, we study the interaction of two differently sized colloids at the channel walls. Recently, we observed that driven large colloids push smaller colloids to the walls. This results in small particle lanes at the walls at early simulation times. In this work, we additionally find that thin lanes are unstable and dissolve over very long time frames. Furthermore, we observe a connection between lane formation and the nonuniform distribution of particles along the channel length. This nonuniform distribution occurs either alongside lane formation or in shared lanes (i.e. lanes consisting of two colloid types).
\end{abstract}

\newcommand{\uproman}[1]{\uppercase\expandafter{\romannumeral#1}}
\newcommand{\lowroman}[1]{\romannumeral#1\relax}
\maketitle

\section{Introduction}

A well-known nonequilibrium phase transition to a segregated steady state is the so-called lane formation. It is induced by a velocity difference of moving individuals. This velocity difference results in collisions or evasive movements between the individuals, eventually causing them to align themselves behind each other (with respect to the flow direction). Lane order phenomena occur over a wide range of length scales. Pedestrians \cite{lee2016,Feliciani2016,Lian2017,Yajima2020}, ants \cite{couzin2003}, propelled microtubules \cite{Memarian2021}, molecular motors \cite{Jose2020} and bacteria \cite{Shimaya2019} are just a few examples.\\
Colloidal systems represent a well-suited model system to study the underlying mechanisms of complex phenomena. For instance, active matter is utilized to model flocks of birds \cite{birdsrome} or other collective behavior \cite{bechinger2019,PSANTON,kaiser2017}. Similarly, passive particles can be used to study the previously mentioned lane formation as well as a phenomenon called band formation \cite{VaterMarc,Vissers2011_band}. Lane formation was found in confined \cite{VaterMarc,Ebrahim2016} and periodic \cite{Dzubiella2002,Chakrabarti2004,Vissers2011,rex2007} channels with oppositely moving colloids. In such systems, hydrodynamic interactions and the influence on the resilience of lane formation was studied in Refs.~\cite{Wysocki2011,Rex2008}. Recently, we found segregation into lanes and a so-called "funneling effect" in confined channels with parallel moving colloids driven by the force of gravity \cite{MarcKay}. This effect describes the interaction of two differently sized colloids in the proximity of a hard wall. Here, the larger particle acts like a funnel and pushes the smaller particle towards the wall. In contrast, binary colloidal mixtures without an applied driving force (i.e., in equilibrium) typically result in larger particles occupying areas near the walls due to entropic depletion forces \cite{paul1998}.\\
Lately, new phenomena have been discovered in colloidal mixtures through increasing the complexity of a binary system by adding a third particle species \cite{toyotama2016,fortini2017,Santos2021}. However, Refs.~\cite{VaterMarc,Vissers2011_band,Ebrahim2016,Dzubiella2002,Chakrabarti2004,Vissers2011,rex2007,MarcKay} studied lane and band formation exclusively in binary systems and to our knowledge, the influence of a third particle species on lane and band formation is unknown, so far.\\
Therefore, we extended our previous system \cite{MarcKay} with a third particle species. The particles are driven by the force of gravity, and all particle species have the same mass density, but differ in their size. Because the gravitational driving force is proportional to the mass of the respective particles, a velocity difference between the particles occurs, enabling the formation of lanes. \\
In this work, we find increasing segregation of the three particle species with increasing driving force, similar to the lane formation observed in binary systems. We study the previously mentioned ”funneling effect”, which we found in \cite{MarcKay}, for both binary and ternary systems in more detail. Here, we find that by observing a larger time frame, the findings of \cite{MarcKay} can be extended. On a larger time scale of the simulations, we find that thin lanes at the channel walls disassemble with time. Finally, we study the band formation in ternary systems. Here, we find that the parameter to detect band formation is increased in some systems which are close to the transition into lanes. In other systems, band formation occurs simultaneously with lane formation.\\
Our work is organized as follows: In Sec.~\ref{sec:model}, we present the numerical model utilized in this work. In Sec.~\ref{sec:observables}, the parameters to detect and quantify lane and band formation are defined. The results and a discussion of the same is given in Sec.~\ref{sec:results}, and finally, we conlude our findings in Sec.~\ref{sec:conclusion}.

\section{Model\label{sec:model}}

We model quasi two-dimensional motion of spherical colloids by employing a Brownian dynamics algorithm, while hydrodynamic interactions are neglected. The stochastical Euler integration of the overdamped Langevin equation provides an update rule for the position of the particle $i$
\begin{align*}
    \mathbf{r}_i (t + \Delta t) = \mathbf{r}_i (t) + \frac{D \Delta t}{k_{\mathrm{B}} T} \mathbf{F}_i \left[ \mathbf{r}_j(t) \right] + \sqrt{2 D \Delta t} \mathbf{R}_i(t) \,,
\end{align*}
with the time-step $\Delta t$, the microscopic diffusion coefficient $D$, the unit of thermal energy $k_\mathrm{B}T$,  and the sum of all forces $\mathbf{F}_i \left[ \mathbf{r}_j(t) \right]$ acting on the $i^\mathrm{th}$ particle, composed of the interactions with all other particles $j$ and the driving force. The components of $\mathbf{R}(t)$ are normally distributed random numbers satisfying the conditions $\bigl \langle R_{k}(t) \bigr \rangle = 0$ and $\bigl \langle R_k (t) R_l (t) \bigr \rangle = \delta_{kl}$. The interaction of two particles $i$ and $j$ with a center-to-center distance $r_{ij}$ is modeled by a Weeks-Chandler-Andersen (WCA) potential
\begin{align*} 
    V(r_{i,j}) = 
    \begin{cases}
    4 \epsilon \left[ \left( \frac{d_{\mathrm{eff}}}{r_{i,j}} \right)^{12} - \left( \frac{d_{\mathrm{eff}}}{r_{i,j}} \right)^{6} \right] + \epsilon & ,\, r \leq \sqrt[6]{2}d_\mathrm{eff} \\
    0 & ,\, r > \sqrt[6]{2}d_\mathrm{eff} \,,
    \end{cases}
\end{align*}
with the effective diameter $d_\mathrm{eff}=(d_i+d_j)/2$, where $d_i$ and $d_j$ are the respective particle diameters. For simplicity, we set the interaction strength to $\epsilon=1$. The initial positions of the particles are randomly distributed within a linear channel, which is characterized by the length $L$ in $x$-direction and width $W$ in $y$-direction. Hard walls, which are implemented by solving the Smoluchowski equation in half space according to Ref.~\cite{hardWall}, confine the particles in $y$-direction and periodic boundaries are employed in $x$-direction. Ternary systems are defined by consisting of three particle types: $n_\mathrm{S}$ small, $n_\mathrm{M}$ medium-sized and $n_\mathrm{L}$ large particles, differing in their diameter. The diameters of the small and large particles are fixed to $d_\mathrm{S}=\sigma$ and $d_\mathrm{L}=2\sigma$, respectively. The diameter of the medium-sized particles is varied in the range $d_\mathrm{M}\in[d_\mathrm{S},d_\mathrm{L}]$, where the two edge cases are equivalent to a binary system with two particle types. The area covered by particle $i$ is calculated by the cross-section area of a sphere $A_\mathrm{part,i}=\pi(d_i/2)^2$. With this, we can calculate the area fraction of a particle species $ty\in\{\mathrm{S},\mathrm{M},\mathrm{L}\}$ with $\alpha_{ty}=(n_{ty}A_{\mathrm{part},ty})/(L W)$, with the number of the particles of said type $n_{ty}$. The total area fraction of all systems is set to $\alpha=\alpha_\mathrm{S}+\alpha_\mathrm{M}+\alpha_\mathrm{L}=0.25$, thus a quarter of the channel area $A_\mathrm{channel}=L\cdot W$ is covered by particles. We studied two types of systems, one where the number of particles of each type is equal $n_\mathrm{S}=n_\mathrm{M}=n_\mathrm{L}$ , and the other where the area fraction of each type is equal $\alpha_\mathrm{S}=\alpha_\mathrm{M}=\alpha_\mathrm{L}$. The former are called \textit{systems with identical particle number} and the latter are called \textit{systems with identical area fraction}. During the simulations, a driving force is applied to the particles in $x$-direction,
\begin{align}
    F_{ty} = F_{\mathrm{S}}\frac{d_{ty}^3}{d_\mathrm{S}^3} \,,\label{eq:mod:Fi}
\end{align}
which models the force of gravity. Here, $F_\mathrm{S}$ is the driving force applied to the small particles, which can be varied freely. This results in the driving force differences
\begin{align}
    \Delta F_\mathrm{MS}&=\frac{d_\mathrm{M}^3-d_\mathrm{S}^3}{d_\mathrm{S}^3}F_\mathrm{S}\,,\,\mathrm{and} \label{eq:DeltaF1}\\ 
    \Delta F_\mathrm{LM}&=\frac{d_\mathrm{L}^3-d_\mathrm{M}^3}{d_\mathrm{S}^3}F_\mathrm{S}\label{eq:DeltaF2}\,.
\end{align}
In this work the diameter of the small particles $\sigma$, the thermal energy $k_\mathrm{B}T$, and the diffusion constant $D$ are set to unity and all other quantities are expressed in terms of these reduced units, such as, e.g., the time with $\tau_{\mathrm{D}} = \sigma^2/D$.

\section{Observables}\label{sec:observables}
We utilized two parameters $\Phi_\text{lane}$ and $\Phi_\text{band}$ to quantify the degree of order in the system. The order parameter $\Phi_\text{lane}$ describes the segregation of the particles in lanes and is chosen similarly to Refs.~\cite{birteullipeter,MarcKay}. To assign a numeric value to the degree of non-uniformity of the particles' $x$-coordinates, the order parameter $\Phi_\text{band}$ is used.

\subsection{Lane formation parameter $\Phi_\mathrm{lane}$}\label{obs:laneformationpara}
\begin{figure}[ht]
    \centering
    \includegraphics[width=0.45\textwidth]{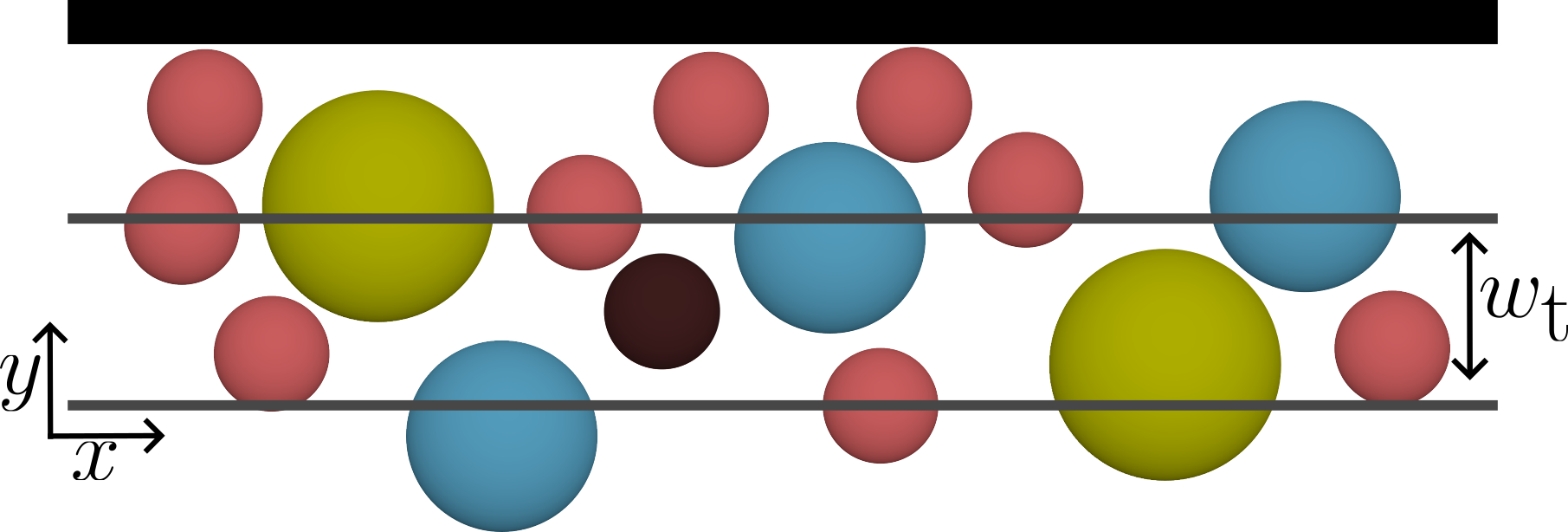}
    \caption{Schematic illustration of the tube used for the calculation of the lane formation parameter $\Phi_\mathrm{lane}$. In the presented example, particle $i$ and the channel walls are represented in black. The tube is depicted with gray lines.}
    \label{fig:obs:tube}
\end{figure}

The order parameter $\Phi_\mathrm{lane}$ is based on Ref.~\cite{MarcKay} and is adapted to describe ternary systems. At a value of $\Phi_\mathrm{lane}=0$ it describes a system where all particles are mixed. When the particles segregate and begin to form lanes, the parameter increases until it approaches its maximum value of $\Phi_\mathrm{lane}=1$, which represents a system where the particles are fully segregated into lanes. In this section, the notation of the particles types is changed to $\mathrm{S}\hat{=}1$, $\mathrm{M}\hat{=}2$, and $\mathrm{L}\hat{=}3$. Let $k,l\in[1,2,3]$ be the particle type of the particles $i$ and $j$. For each time step, the parameter $\phi_{\mathrm{lane,}i}$ is calculated for particle $i$ by defining a tube of width $w_{\mathrm{t}}=\rho^{1/2}$ and counting each other particle $j$ inside (see Fig.~\ref{fig:obs:tube}). Here, $\rho$ denotes the total particle density. For every particle $j$ inside the tube, either $n_{i,+}$ ($k=l$) or $n_{i,-}$ ($k\neq l$) is increased. By how much these are increased depends on the total number of particles of the respective types $k$ and $l$. To determine this, we introduce the vectors $\textbf{I}$ and $\textbf{J}$. The components $I_\mathrm{k}=J_\mathrm{l}=1$ and all other are zero. For a binary system, the increase values $\hat{f}^\mathrm{bin}_{k,l}$ are given by 
\begin{align*}
    \hat{f}^\text{bin} = \begin{pmatrix}
                            1 & \frac{n_\text{S}}{n_\text{M}} \\
                            \frac{n_\text{M}}{n_\text{S}} & 1    
                         \end{pmatrix}\,. 
\end{align*}
For a ternary system with three particle groups, the increase values $\hat{f}^\mathrm{ter}_\text{k,l}$ are defined by 
\begin{align*}
    \hat{f}^\text{ter} = \begin{pmatrix}
        1 & \frac{n_\text{S}}{2n_\text{M}} & \frac{n_\text{S}}{2n_\text{L}} \\
        \frac{n_\text{M}}{2n_\text{S}} & 1 & \frac{n_\text{M}}{2n_\text{L}} \\
        \frac{n_\text{L}}{2n_\text{S}} & \frac{n_\text{L}}{2n_\text{M}} & 1 
    \end{pmatrix}\,. 
\end{align*}
The respective value to increase either $n_{i,+}$ or $n_{i,-}$ is determined using
\begin{align*}
    \textbf{I}^T\hat{f}^\mathrm{bin/ter}\textbf{J}\,.
\end{align*}
As an example, let particle $i$ be a small particle ($k=1$). In the case that the center of the particle $j$ is inside the tube and is also a small particle ($l=1$), we increase $n_\mathrm{i,+}$ by 
\begin{align*}
     (1,0,0)\hat{f}^\mathrm{ter}(1,0,0)^\intercal=1\quad .
\end{align*}
In another case, where the particle $j$ is a large particle ($l=3$), we increase $n_\mathrm{i,-}$ by
\begin{align*}
     (1,0,0)\hat{f}^\mathrm{ter}(0,0,1)^\intercal=\frac{n_\mathrm{L}}{2n_\mathrm{S}}\,.
\end{align*}
By means of these results, the order parameter of the $i^\mathrm{th}$ particle can be calculated by
\begin{align*}
    \phi_{\mathrm{lane},i} = \left| \frac{n_{i,+} - n_{i,-}}{n_{i,+} + n_{i,-}} \right| \,.
\end{align*}
This procedure is repeated for every particle $i$ and the results are averaged to receive the lane formation parameter
\begin{align*}
    \Phi_{\mathrm{lane}} = \frac{1}{N} \sum _{i=1} ^{N} \phi_{\mathrm{lane},i} \,,
\end{align*}
with the number of particles $N$. In binary systems, the lane formation parameter for both particle types is by definition equal, thus it is averaged over all particles, regardless of the type. This is different in ternary systems, where the parameter may vary between particle types and is therefore averaged over all types separately
\begin{align*}
    \Phi_{\mathrm{lane},ty} = \frac{1}{n_{ty}} \sum _{i=1} ^{n_{ty}} \phi_{\mathrm{lane},i} \,.
\end{align*}

\subsection{Band formation parameter $\Phi_\mathrm{band}$}
To determine the order parameter $\Phi_\mathrm{band}$, analogously to Ref.~\cite{VaterMarc}, the channel is divided into $n_\mathrm{bin}=72$ bins in $x$-direction and a histogram for the $x$-coordinates of the particles is created. The mean number of particles $M$ in each bin is calculated by 
\begin{align*}
    M=\frac{N}{n_\mathrm{bin}}
\end{align*}
and the variance by
\begin{align*}
    S^2=\sum_{i=1}^{n_\mathrm{bin}}\frac{(n_i-M)^2}{n_\mathrm{bin}}\,.
\end{align*}
with the number of particles $n_i$ in bin $i$. The ratio between the variance and the squared mean number of particles is the band parameter 
\begin{align*}
    \Phi_\mathrm{band}=\frac{S^2}{M^2}\,.
\end{align*}
It vanishes if the particles are distributed uniformly and $\Phi_\mathrm{band} = 1$ is reached in a state corresponding to a half empty channel, with evenly distributed particles in the other half. The maximum value of $\Phi_\mathrm{band} = n_\mathrm{bin}$ can only be reached if all particles accumulate in a common bin. In this work, the band formation parameter is calculated separately for each particle type in order to detect density inhomogeneities of a single particle type.

\subsection{Lane detection}
\begin{figure}[ht]
    \centering
    \includegraphics[width=0.45\textwidth]{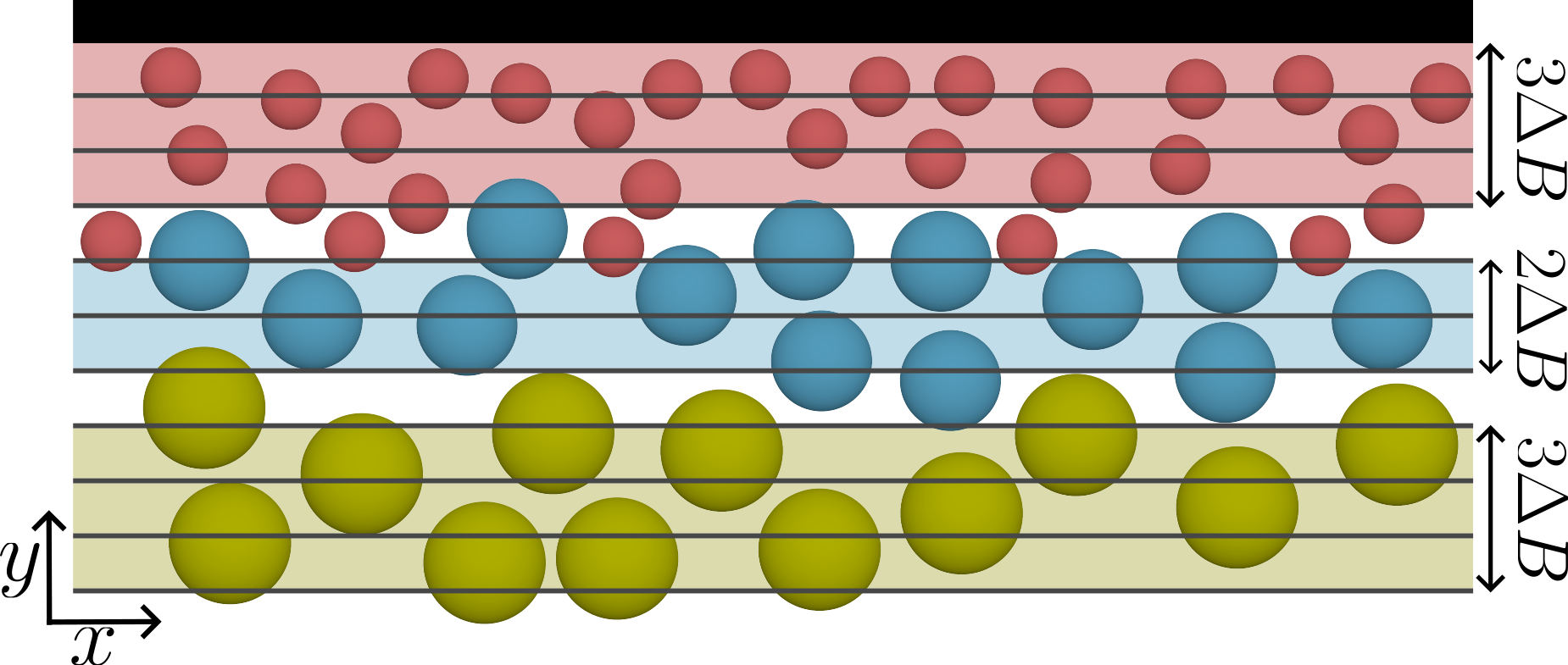}
    \caption{Schematic illustration of the lane detection. The hard wall is shown as a black bar. The bins of the histogram are depicted with gray lines. The bins shaded with a color are assigned to the respective particle species. The unshaded bins are unassigned.}
    \label{fig:obs:lanedet}
\end{figure}

While the lane formation parameter assigns a numerical value to the degree of lane formation in the system, it contains no information about the number and width of lanes. To determine this, as well as which particle type forms lanes closest to the walls, a histogram over the $y$-coordinates of the particles is computed. The channel is divided into bins with width $\Delta B=1$, resulting in the number of bins $n_\mathrm{bin}=W/\Delta B$. The area fraction of particle type $ty\in\{S,M,L\}$ in bin $v$ is given by
\begin{align}
    \alpha_{ty,v}=\frac{n_{ty,v}\cdot A_ty}{L\cdot \Delta B}\,.
\end{align}
with $n_{ty,v}$ the number of particles in bin $v$ and $A_{ty}$ the area of a particle of particle type $ty$. Now, bin $v$ is assigned to the particle type with the highest area fraction in this specific bin (see Fig.~\ref{fig:obs:lanedet}), but only if it surpasses the area fraction of the respective particle type of the whole channel $\alpha_{ty,v} > \alpha_{ty}$. Otherwise, the bin is not assigned to any particle type. A sequence of bins belonging uninterruptedly to the same particle type is defined as a lane. Bins belonging to a different particle type interrupt lanes, while unassigned bins do not. Thus, the width of a lane can be calculated by the number of bins of an uninterrupted sequence and the bin width $\Delta B$.

\section{Results and Discussion\label{sec:results}}
In this study, all channel lengths are fixed to $W=50$ in $y$-direction and $L=400$ in $x$-direction. We justify this specific length $L$, since  we consider it a reasonable compromise between computational effort and the limitation of finite size effects due to periodic boundary conditions (for a short discussion on this, see Appendix A). Each simulation is split into two parts. The first part consists of $1.5\cdot 10^{8}$ time steps and the second part of $1.5\cdot 10^{6}$ time steps with a step size of $\Delta t=10^{-5}$. The simulation duration of the first part is chosen such that the lane parameter is allowed to reach a stable value and in excess of experimentally viable time frames (see Appendix B). The order parameters $\Phi_\mathrm{lane}$ and $\Phi_\mathrm{band}$ are analyzed over the second run. To obtain sufficient statistics, each simulation is run 20 times with different initial particle positions and different random numbers. In Appendix D, we roughly estimate experimentally viable observation times to $T_\mathrm{exp}\approx\SI{0.5}{\hour}$ with experimental values from Refs.~\cite{Siems2012,viswater,Alsaadawi2021}. The maximum simulation time $T_\mathrm{sim}=1500\,\tau_\mathrm{D}$ of this work translates to $T_\mathrm{sim}\approx\SI{63}{\hour}$ for the chosen values. With $T_\mathrm{sim}\gg T_\mathrm{exp}$, we can safely assume our simulation duration to be sufficient.

\subsection{Introduction to lane formation and critical force}
\begin{figure}[ht]
    \centering
    \includegraphics[width=0.5\textwidth]{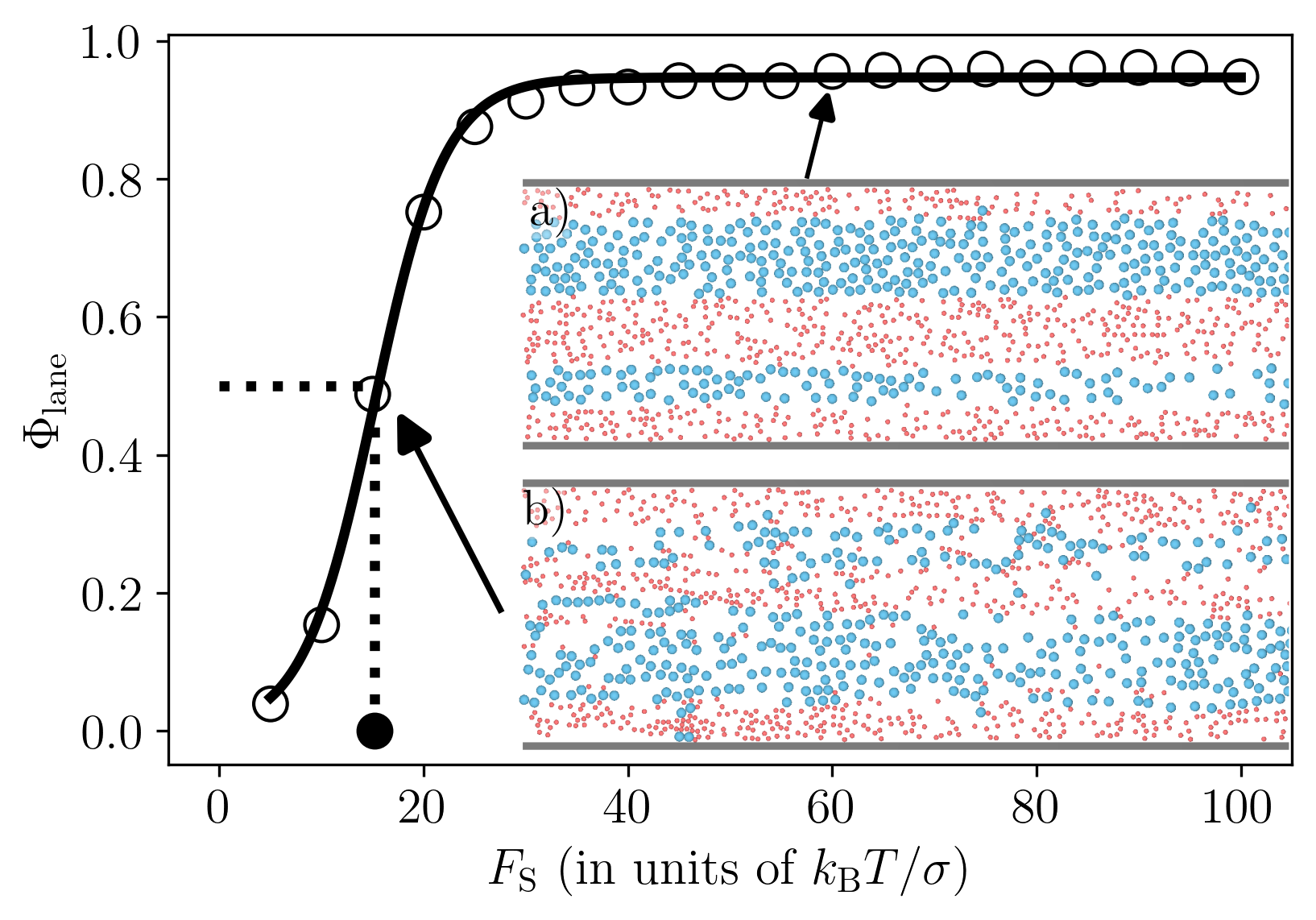}
    \caption{Lane formation in a binary system with width $W=50$ and length $L=400$. At high driving forces of a) $F_\mathrm{S}=60$ the particles are fully demixed and clear lanes are visible. At low driving forces of b) $F_\mathrm{S}=15$ the system is in the transition to lane formation, but lanes are not yet fully established. The solid line shows the fit function $\Phi_\mathrm{lane,bin}$. The dashed lines mark the value $\Phi_\mathrm{lane}=0.5$ and the corresponding critical force $F_\mathrm{crit}=15.2$.}
    \label{fig:res:binlanepics}
\end{figure}

This section acts as an introduction to the methodology used in this work. Additionally, it serves to be more easily be able to directly compare lane formation in binary and ternary systems.\\
All particles in this work are subject to a force, according to Eq.~\eqref{eq:mod:Fi}, which drives them through the channel in dependence of their diameter. This difference in the particle driving forces induces collisions, which can eventually result in the demixing of the particle types perpendicular to the movement direction. Fig~\ref{fig:res:binlanepics} shows the lane formation parameter $\Phi_\mathrm{lane}$ in dependence of the driving force $F_\mathrm{S}$ as well as two inset figures with exemplary channel configurations. Fig.~\ref{fig:res:binlanepics}~a) depicts a section of a binary system with $n_\mathrm{S}=2122$ small and $n_\mathrm{L}=1061$ large particles, with a driving force of $F_\mathrm{S}=60$ for the small and $F_\mathrm{L}=480$ for the large particles. Fig.~\ref{fig:res:binlanepics}~b) shows the same channel at the driving force $F_\mathrm{S}=15$. This binary system represents an case ($d_\mathrm{S}=d_\mathrm{M}=1.0$) of a ternary \textit{systems with identical particle number} with twice as many small/medium-sized particles than large particles. At this driving force, a separation perpendicular to the driving force is already apparent. However, the interfaces between two neighboring lanes are not fully established. In other words, some particles move through lanes of the other particle type. By further increasing the driving forces to $F_\mathrm{S}=60$ and $F_\mathrm{L}=480$ as depicted in Fig.~\ref{fig:res:binlanepics}~a), the particle types are virtually fully demixed and well-defined interfaces between lanes exist.\\
The lane formation order parameter $\Phi_\mathrm{lane}$ described in Sec.~\ref{sec:observables} assigns a numerical value to each system, which represents the degree of lane formation. Fig.~\ref{fig:res:binlanepics} shows the lane formation parameter for the driving forces $F_\mathrm{S}\in[5,100]$. With increasing driving force, the lane formation order parameter increases until it reaches a plateau just below $\Phi_\mathrm{lane}=1.0$ at $F_\mathrm{S}=30$. Since for each studied parameter set, the lane formation order parameter follows a sigmoidal shape, we use the following fit function to represent these data
\begin{align}\label{eq:res:lanefitBin}
     \Phi_\mathrm{lane,bin}(F)=\frac{1}{4}\Bigg(\left(\mathrm{cos}(a)+1\right)\biggl(\mathrm{tanh}\Bigl(\frac{F-b}{c}\Bigr)+1\biggr)\Bigg)\,,
\end{align}
with $a$, $b$ and $c$ the fit parameters. This fit function is adapted from Ref.~\cite{VaterMarc}, where the $\mathrm{cos}$ term keeps the fit between 0 and 1 and the $\mathrm{tanh}$ term provides the sigmoidal shape. The critical force $F_\mathrm{crit}$ is defined as the driving force at which the fit reaches $\Phi_\mathrm{lane,crit}=0.5$. For this binary system, the critical force is $F_\mathrm{crit}=15.2$, which is marked by the dashed lines and black dot in Fig.~\ref{fig:res:binlanepics}.

\subsection{Lane formation and critical force in ternary systems}
Similarly to the lane formation in binary systems, we observe that a difference in the driving force between particle types results in the formation of lanes in ternary systems, as well. However, by adding a third particle type, differences to the process of lane formation in binary systems arise. Since we find that there are almost no qualitative differences between \textit{systems with identical particle number} and \textit{systems with identical area fraction} systems, we discuss the lane formation based on the former system type. A comparison between system types is given in Appendix C. Fig.~\ref{fig:res:terlanepics140} depicts three snapshots of a ternary system with a medium-sized particle diameter $d_\mathrm{M}=1.40$ and the dependency of the lane formation parameter $\Phi_\mathrm{lane}$ on the driving force $F_\mathrm{S}$.
At a low driving force of $F_\mathrm{S}=20$ the separation of any particle type perpendicular to the driving force into lanes is not yet developed, which can be seen in Fig.~\ref{fig:res:terlanepics140}~a). By increasing the driving force to $F_\mathrm{S}=50$, as presented in Fig.~\ref{fig:res:terlanepics140}~b), the largest particles segregate from the other two particle types and form lanes. The smallest and medium-sized particles are still mixed and share lanes, we will from now on refer to such lanes as shared lanes. Fig.~\ref{fig:res:terlanepics140}~c) shows a channel with the driving force $F_\mathrm{S}=100$. Here, all particle types form lanes, but the lanes of the small and medium-sized particles are not fully established. We suspect however that increasing the driving force even more would further segregate the small and medium-sized particles. However, simulations with the chosen timestep of $\Delta t=10^{-5}$ for driving forces exeeding $F_\mathrm{S}=100$ were unstable. In Fig.~\ref{fig:res:terlanepics140}~d), the order parameter $\Phi_\mathrm{lane}$ in dependence on the driving force $F_\mathrm{S}$ can be seen. In contrast to the binary system, the lane formation order parameter $\Phi_\mathrm{lane}$ is calculated for all particle types separately. At low driving forces, all particle types stay mixed at $\Phi_\mathrm{lane}=0$. By increasing the driving force, the order parameter of the largest particles $\Phi_\mathrm{lane,L}$ increases rapidly until it saturates just below $\Phi_\mathrm{lane}=1$. The qualitative dependency appears very similar to the order parameter $\Phi_\mathrm{lane}$ in binary systems (compare Fig.~\ref{fig:res:binlanepics}). The small and medium-sized particles behave differently. The courses of the parameters $\Phi_\mathrm{lane,S}$ and $\Phi_\mathrm{lane,M}$ show a double sigmoidal shape, which can be understood with our definition of $\hat{f}^\mathrm{ter}$. The order parameter of the small and medium sized particles increases with the segregation of the large particles, even though the small and medium sized particles still coexist in the same shared lane. This is however intended behavior, since we want $\Phi_\mathrm{lane}$ to vanish in the case that we have the same particle distribution as in the equilibrium system. Only when the small and medium sized particles reach a state where they occupy their own lanes should their order paremeters approach unity. In the range of $F_\mathrm{S}\in[40,60]$, the lane formation parameter for the small and medium-sized particles reaches a plateau. In this plateau region, shared lanes exist. At higher driving forces, the small and medium-sized particles also demix and the corresponding order parameter $\Phi_\mathrm{lane}$ increases. It is reasonable to assume that for even higher driving forces, the order parameter of the small and medium-sized particles would also asymptotically approach $\Phi_\mathrm{lane}=1$. The existence of shared lanes in ternary systems requires an adjustment of the binary fit function Eq.~\eqref{eq:res:lanefitBin}. Thus, a second term is added to provide a suitable fit function
\begin{align}\label{eq:res:lanefitTer}
     \Phi_\mathrm{lane,ter}(F)=\frac{1}{8}\Bigg(&\left(\mathrm{cos}(a_1)+1\right)\biggl(\mathrm{tanh}\Bigl(\frac{F-b_1}{c_1}\Bigr)+1\biggr)\\
    +&\left(\mathrm{cos}(a_2)+1\right)\biggl(\mathrm{tanh}\Bigl(\frac{F-b_2}{c_2}\Bigr)+1\biggr)\Bigg)\,,\nonumber
\end{align}
which results in a double sigmoidal shape as depicted in Fig.~\ref{fig:res:terlanepics140}~d). Analogously to the binary systems, the critical force $F_\mathrm{crit}$ is defined as the driving force at which the respective fit exceeds $\Phi_\mathrm{lane,crit}=0.5$. For the system depicted in Fig.~\ref{fig:res:terlanepics140}~d) the critical forces are $F_\mathrm{crit,S}=88.7$, $F_\mathrm{crit,M}=82.2$ and $F_\mathrm{crit,L}=22.4$.
\begin{figure*}[t]
     \centering
     \includegraphics[width=\textwidth]{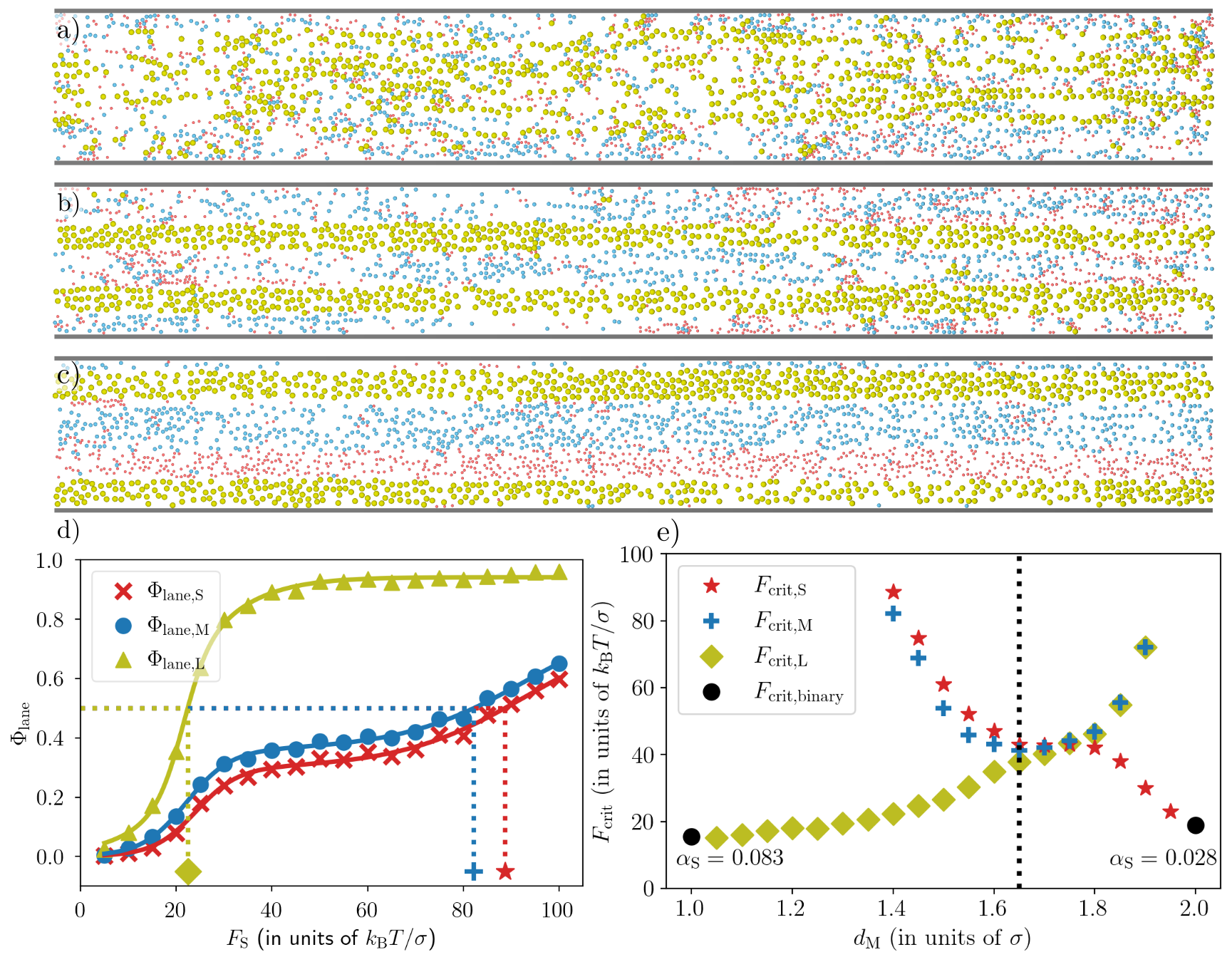}
     \caption{Lane formation in a ternary system. Fig.~a), c), d) and e) show the dependence of the lane formation order parameter $\Phi_\mathrm{lane}$ on the driving force for a medium-sized particle diameter $d_\mathrm{M}=1.40$. For low driving forces of a) $F_\mathrm{S}=20$ the particles of each type start to form short chains. By increasing the driving force to b) $F_\mathrm{S}=50$ the largest particles formed lanes, while the small and medium-sized particles move in shared lanes. For the driving force of c) $F_\mathrm{S}=100$ distinct lanes form. The lanes of the largest particles are fully separated. The small and medium-sized particle would further demix by increasing the driving force. Fig.~e) depicts the dependence of the critical force of each particle type on the medium-sized particle diameter $d_\mathrm{M}$. The black circles correspond to binary systems. The dotted black line marks $d_\mathrm{M}=1.65$.}
     \label{fig:res:terlanepics140}
\end{figure*}

\subsubsection{Dependence on the medium-sized particle diameter}
\begin{figure}[t]
     \centering
     \includegraphics[width=0.5\textwidth]{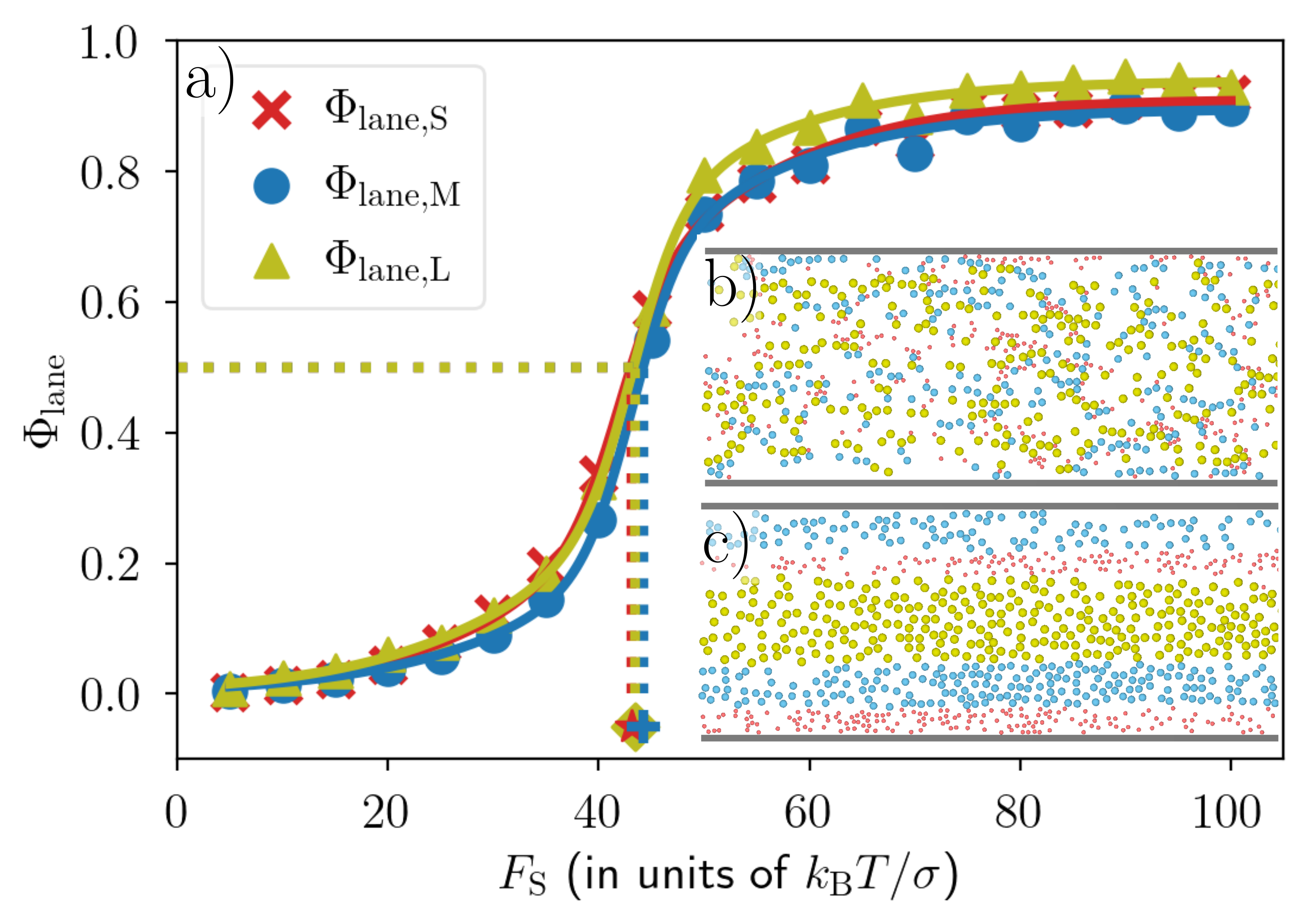}
     \caption{Lane formation in ternary systems with a medium-sized particle diameter $d_\mathrm{M}=1.75$. While at a low driving force of b) $F_\mathrm{S}=20$ the channel can be found in an unordered state, a high driving force of c) $F_\mathrm{S}=90$ results in the segregation of the different particle types.}
     \label{fig:res:terlanepics175}
\end{figure}

To better understand the size effect of the medium-sized particles on lane formation, we varied the medium-sized particle diameter in the range $d_\mathrm{M} \in [d_\mathrm{S}, d_\mathrm{L}]$. We present the critical forces for these systems in Fig.~\ref{fig:res:terlanepics140}~e). Both edge cases of $d_\mathrm{M}=d_\mathrm{S}$ and $d_\mathrm{M}=d_\mathrm{L}$ (marked with a black circle) correspond to binary systems with the small particle area fraction $\alpha_\mathrm{S}=0.083$, and 0.028 (twice or half as many small than large particles). The critical forces are calculated by means of the binary and ternary fit funtions Eqs.~\eqref{eq:res:lanefitBin} and \eqref{eq:res:lanefitTer}. The absence of data points (e.g. for the small and medium-sized particles below $d_\mathrm{M}=1.40$) indicate that the lane formation parameter of the respective particle types did not surpass the critical value $\Phi_\mathrm{lane,crit}=0.5$.\\
By increasing the diameter of the medium-sized particles, the large particle critical force $F_\mathrm{crit,L}$ rises. The critical force of the small $F_\mathrm{crit,S}$ and medium-sized $F_\mathrm{crit,M}$ particles decreases up to $d_\mathrm{M}=1.65$. At this point, we observe a minimum for the medium-sized particle critical force and a saddle point for the critical force of the small particles. This specific dependency can be explained by the driving force differences between the particle types in Eq.~\eqref{eq:DeltaF1} and Eq.~\eqref{eq:DeltaF2}. If we assume that a larger size difference results in a smaller critical force and vice versa, we can calculate the medium-sized particle diameter $d_\mathrm{M}$ at which the critical force of the medium-sized particles $F_\mathrm{crit,M}$ is minimal. At this point
\begin{align*}
     \Delta F_\mathrm{MS}&=\Delta F_\mathrm{LM}\\
     d_\mathrm{M}^3F_\mathrm{S}-d_\mathrm{S}^3F_\mathrm{S}&=d_\mathrm{L}^3F_\mathrm{S}-d_\mathrm{M}^3F_\mathrm{S}\\
     d_\mathrm{M,min}&=\sqrt[3]{\frac{d_\mathrm{L}^3+d_\mathrm{S}^3}{2}}
\end{align*}
holds, because varying the medium-sized particle diameter would require a larger driving force to segregate the medium-sized particles from either the small or the large particles. With the small and large particle diameters are set to $d_\mathrm{S}=1.0$ and $d_\mathrm{L}=2.0$, we obtain $d_\mathrm{M,min}=\sqrt[3]{9/2}\approx 1.65$ (marked by the dotted black line). The results in Fig.~\ref{fig:res:terlanepics140}~e) confirm this specific value with a small deviation, which could, e.g., be caused by size effects of the particles. We find that the critical forces of all particle types are virtually the same at around $d_\mathrm{M}=1.75$. In Fig.~\ref{fig:res:terlanepics175}~a) the dependence of the lane formation parameter of each particle type on the driving force $F_\mathrm{S}$ is depicted for $d_\mathrm{M}=1.75$. The critical forces are $F_\mathrm{crit,S}=43.1$, $F_\mathrm{crit,M}=44.2$ and $F_\mathrm{crit,L}=43.4$. By comparing Fig.~\ref{fig:res:terlanepics140}~d) with Fig.~\ref{fig:res:terlanepics175}~a), the influence of the medium-sized particle diameter is apparent. At low driving forces $F_\mathrm{S}=20$, both systems are mixed, as can be seen in Fig.~\ref{fig:res:terlanepics140}~a) and \ref{fig:res:terlanepics175}~b). However, we observe for $d_\mathrm{M}=1.75$ that the ordering process of each particle type occurs at the same driving force $F_\mathrm{S}$ until they are fully demixed as presented in Fig.~\ref{fig:res:terlanepics175}~c). Consequently, the double sigmoidal shape vanishes and the course of each lane formation parameter overlaps and resembles that of a binary system, as depicted in Fig.~\ref{fig:res:binlanepics}.
\subsection{Funneling effect at the channel walls on long timescales}
\begin{figure}[ht]
    \centering
    \includegraphics[width=0.45\textwidth]{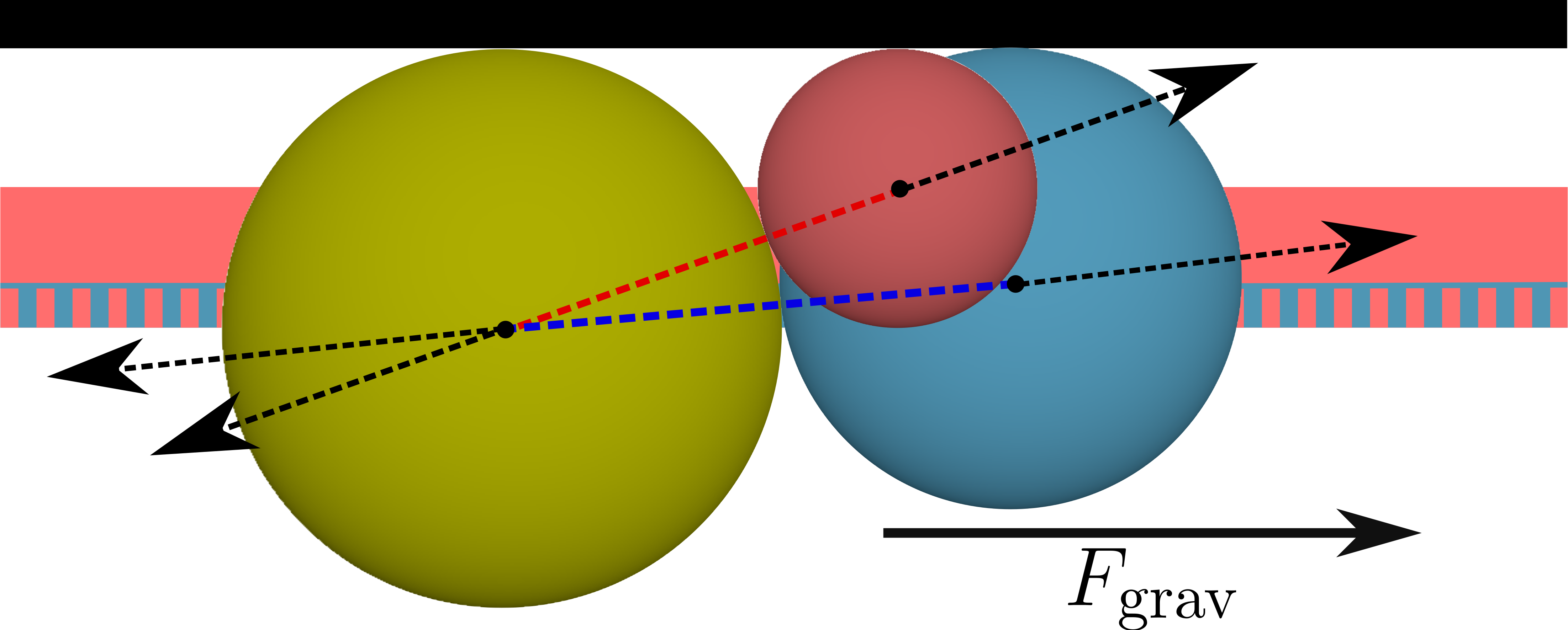}
    \caption{Schematic illustration of the funneling effect at hard walls. The arrow $F_\mathrm{grav}$ shows the drive direction of the particles. The presented particles differ in their size by large $d_\mathrm{L}=2.0$ (yellow), medium-sized $d_\mathrm{M}=1.65$ (blue) and small $d_\mathrm{S}=1.0$ (red). The solid black arrow denotes the driving force of the particles. The dotted red and blue lines connect the particle center.}
    \label{fig:res:funnelPic}
\end{figure}

Recently, we found a nonequilibrium phenomenon in binary systems with small and large particles, which we called the funneling effect \cite{MarcKay}. Here, smaller particles can squeeze between larger particles and a hard wall, which results mostly in boundary lanes (i.e., the lanes closest to the channel walls) of small particles. This effect is illustrated in Fig.~\ref{fig:res:funnelPic}. The width of the red area represents the area that a small (red) particle can enter but a large (yellow) particle cannot, we will refer to this area as funnel area from now on. A collision between the large and the small particle results in the displacement along the dashed black lines extending from the red dashed line. If the center of the small particle is inside the respective funnel area, a collision with a large particle will always push the small particle to the wall. This mechanism works analogously for a large and medium-sized particle (see dashed blue area) or a medium-sized and small particle. This lead to the observation that the boundary lanes in these systems mainly consisted of small particles.

\subsubsection{Binary systems}
\begin{figure}[ht]
    \centering
    \includegraphics[width=.45\textwidth]{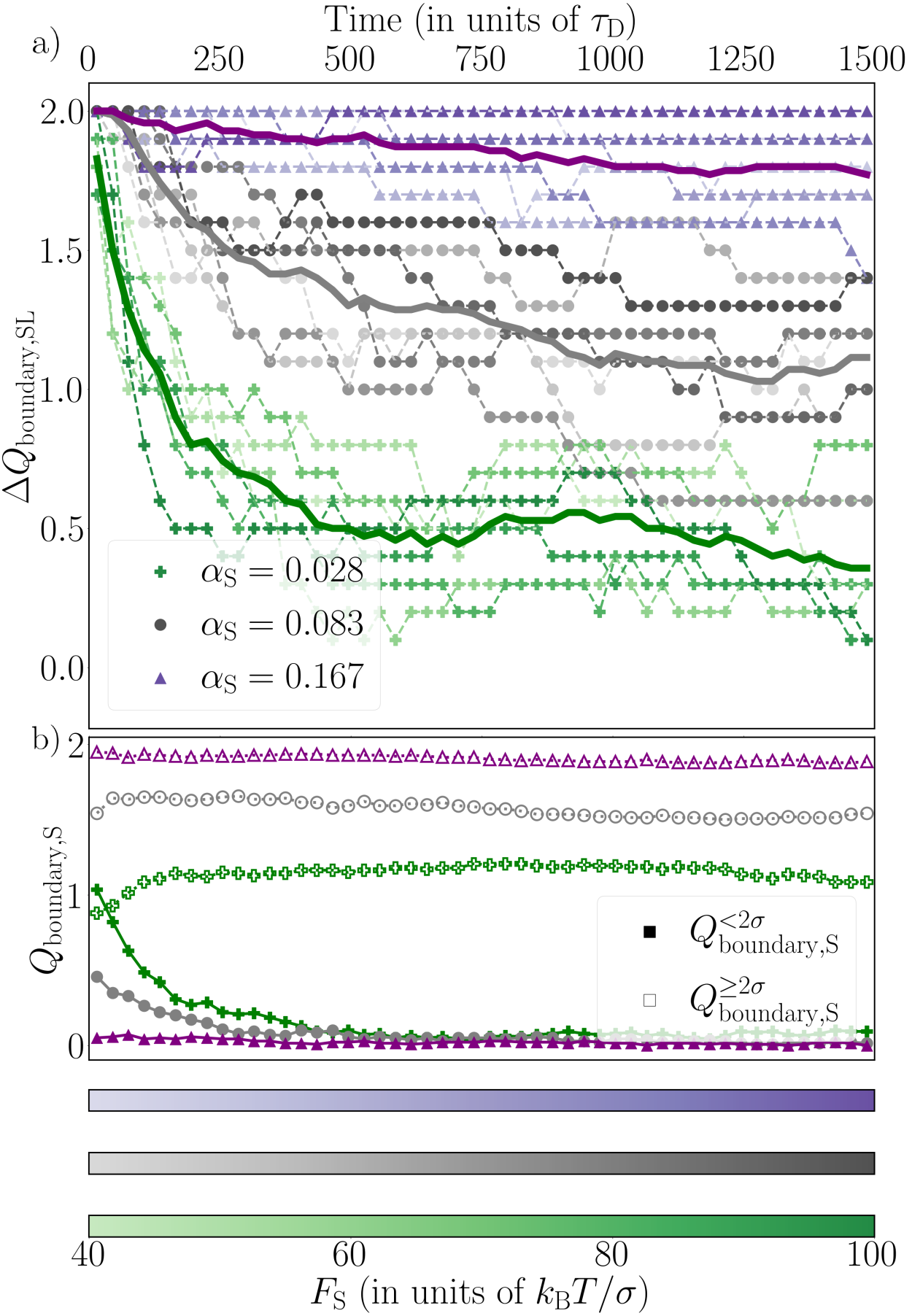}
    \caption{a) Difference in the number of small and large particle boundary lanes $\Delta Q_\mathrm{boundary,SL}$ and b) number of thin $Q^{<2\sigma}_\mathrm{boundary,S}$ and thick $Q^{\geq 2\sigma}_\mathrm{boundary,S}$ small particle boundary lanes in dependence of the simulation time $t$ for the driving forces $F_\mathrm{S}\in[40,100]$ and the area fractions $\alpha_\mathrm{S}=0.028$, 0.083 and 0.167 in binary systems.}
    \label{fig:res:NumberEdgeOverTS_bin}
\end{figure}

\begin{figure}[h]
    \centering
    \includegraphics[width=0.45\textwidth]{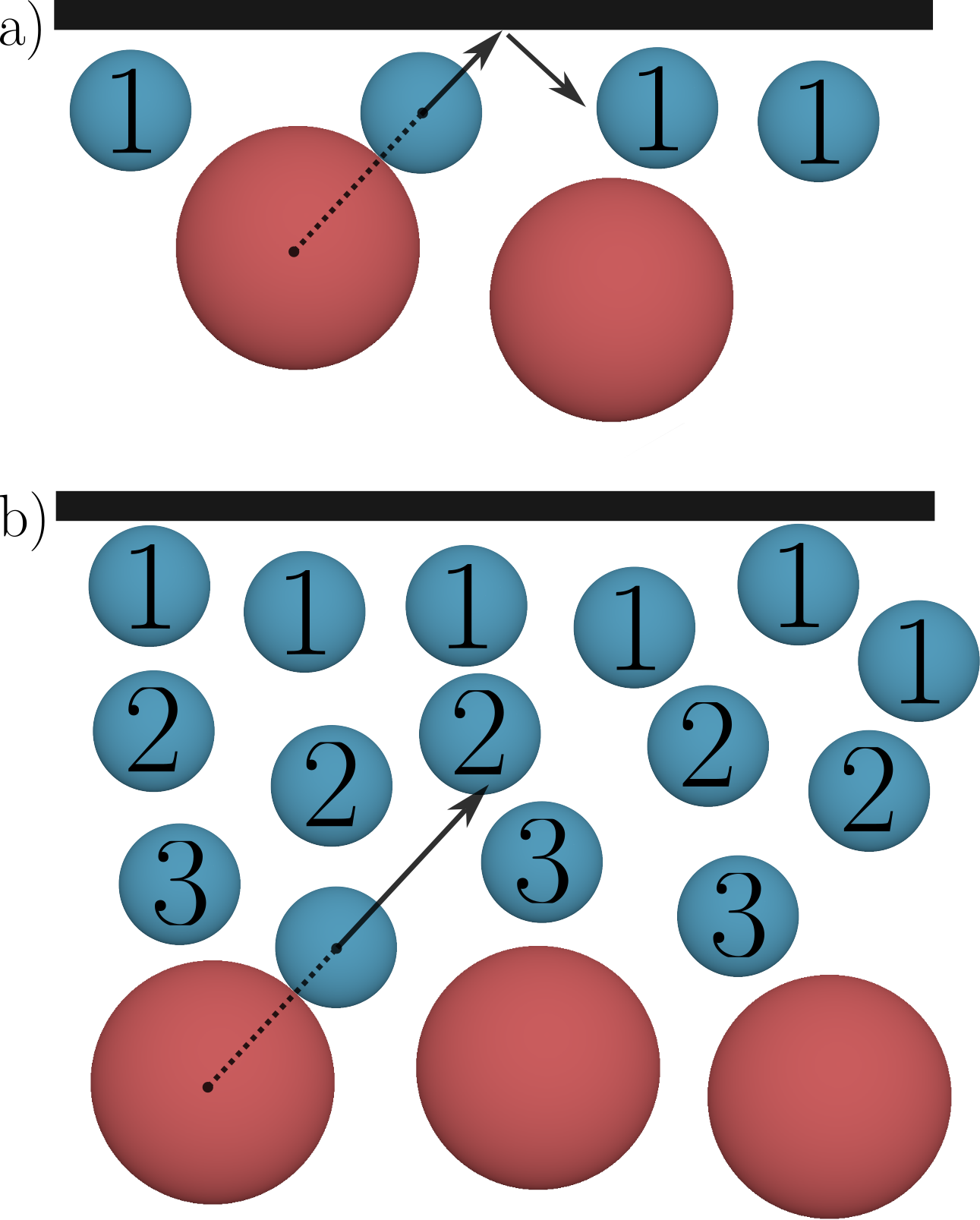}
    \caption{Schematic illustration of two possible funnel events. A boundary lane particle which is located directly at the wall is referred to as particle 1, a boundary lane particle which is located some distance away from the wall as particle 2, and a boundary lane particle that is positioned at the interface to the next lane of larger, faster particles, as particle 3. Picture a) shows a thin boundary lane of small particles. Here, the collision between the large and small particle could result in the reflection of the small particle at the wall. Picture b) depicts a thick boundary lane of small particles. In this case, the small particle can transfer its momentum in $y$-direction to other particles in the boundary lane and occupy open area between them.}
    \label{fig:res:bounceback}
\end{figure}

In order to check whether boundary lanes in binary systems also consist mainly of the smallest particles on long time scales, we increase the simulation duration from $100\,\tau_\mathrm{D}$ in Ref.~\cite{MarcKay} to $1500\,\tau_\mathrm{D}$. Additionally, we studied different area fractions $\alpha_\mathrm{S}=0.028$, 0.083 and 0.167 (which correspond to a ratio of the area fractions $\alpha_\mathrm{S}/\alpha_\mathrm{L} = 1/8$, $1/2$ and $2$).\\
Fig.~\ref{fig:res:NumberEdgeOverTS_bin}~a) presents the difference between the number of small and large particle boundary lanes $\Delta Q_\mathrm{boundary,SL}=Q_\mathrm{boundary,S}-Q_\mathrm{boundary,L}$ for the driving forces $F_\mathrm{S}\in[40,100]$ over time. A value of $\Delta Q_\mathrm{boundary,SL}=2.0$ can only be reached, when every boundary lane consists of small particles, and the difference vanishes if boundary lanes of both particle types occur equally often. In agreement with Ref.~\cite{MarcKay}, we observe mainly boundary lanes of small particles at early simulation times $t=15\,\tau_\mathrm{D}$, regardless of the driving force $F_\mathrm{S}$ or small particle area fraction $\alpha_\mathrm{S}$. With increasing simulation duration, some small particle boundary lanes are displaced by large particle boundary lanes. The lower the area fraction $\alpha_\mathrm{S}$ is, the more boundary lanes of small particles are displaced. This can be explained by the following: The funneling effect only influences particles which are in proximity of a wall, specifically whose center is positioned in the funnel area. Since at the start of the simulation, the particles are randomly distributed, the amount of small particles located inside the funnel area is higher for a greater area fraction $\alpha_\mathrm{S}$. Consequently, the widths of these early small particle boundary lanes, induced by the funneling effect, depends on the area fraction $\alpha_\mathrm{S}$. Fig.~\ref{fig:res:NumberEdgeOverTS_bin}~b) shows the amount of thin (i.e. lane width $W_\mathrm{boundary,S}<2$) and thick (i.e. lane width $W_\mathrm{boundary,S}\geq 2$) boundary lanes of small particles. At early simulation times $t=15\,\tau_\mathrm{D}$, we observe a higher amount of thin small particle boundary lanes for the area fractions $\alpha_\mathrm{S}=0.028$ and 0.083. With time, these thin boundary lanes vanish, while the respective amount of thick boundary lanes stays roughly constant. Thus, the thin small particle boundary lanes do not become wider with time, but are rather displaced by lanes of large particles. Hence, the instability of thin boundary lanes consiting of small particles explains their decrease in Fig.~\ref{fig:res:NumberEdgeOverTS_bin}~a). In the following, we present a possible explanation to understand this instability of thin boundary lanes. We consider a lane of small particles, that have accumulated at the wall due to the funneling effect early on in the simulations. We refer to a boundary lane particle which is located directly at the wall as particle 1, a boundary lane particle which is located some distance away from the wall as particle 2, and a boundary lane particle that is positioned at the interface to the next lane of larger, faster particles as particle 3. Now we distinguish between two scenarios.\\
In the first one (see Fig.~\ref{fig:res:bounceback}~a)), we consider a boundary lane with a width equivalent to the diameter of a single small particle, denoted as $d_\mathrm{S}$. This boundary lane contains virtually no particle 2, and particle 1 and 3 are identical. In this situation, particle 1 can diffuse on long timescales into the lane of faster particles, which destabilizes the boundary lane long term. Additionally, in case of a funnel event (large particle pushes small particle outwards, which is illustrated in Fig.~\ref{fig:res:bounceback}~a) and b)), the only possibility for particle 1 is to move towards the wall and to be reflected by it.\\
In the second scenario (see Fig.~\ref{fig:res:bounceback}~b)), we consider a thicker boundary lane of small particles. Now, mainly particle 3 can enter the lane of faster particles by diffusion in $y$-direction. In this scenario, a different possibility to react to a funnel event exists for particle 3. Particle 3 can transfer its momentum in $y$-direction onto particles 1 and 2. In addition, particle 3 can occupy the packing induced empty areas between two particles, a possibility which does not exist in the first scenario. As a result, the structure becomes more tightly meshed. In a sense, the funnel event in this scenario can be considered as a collision with a "soft" wall. Those effects could lead to a better stabilization of the boundary lanes in the second scenario in comparison to the first scenario.

\subsubsection{Ternary systems}
\begin{figure}[ht]
    \centering
    \includegraphics[width=.45\textwidth]{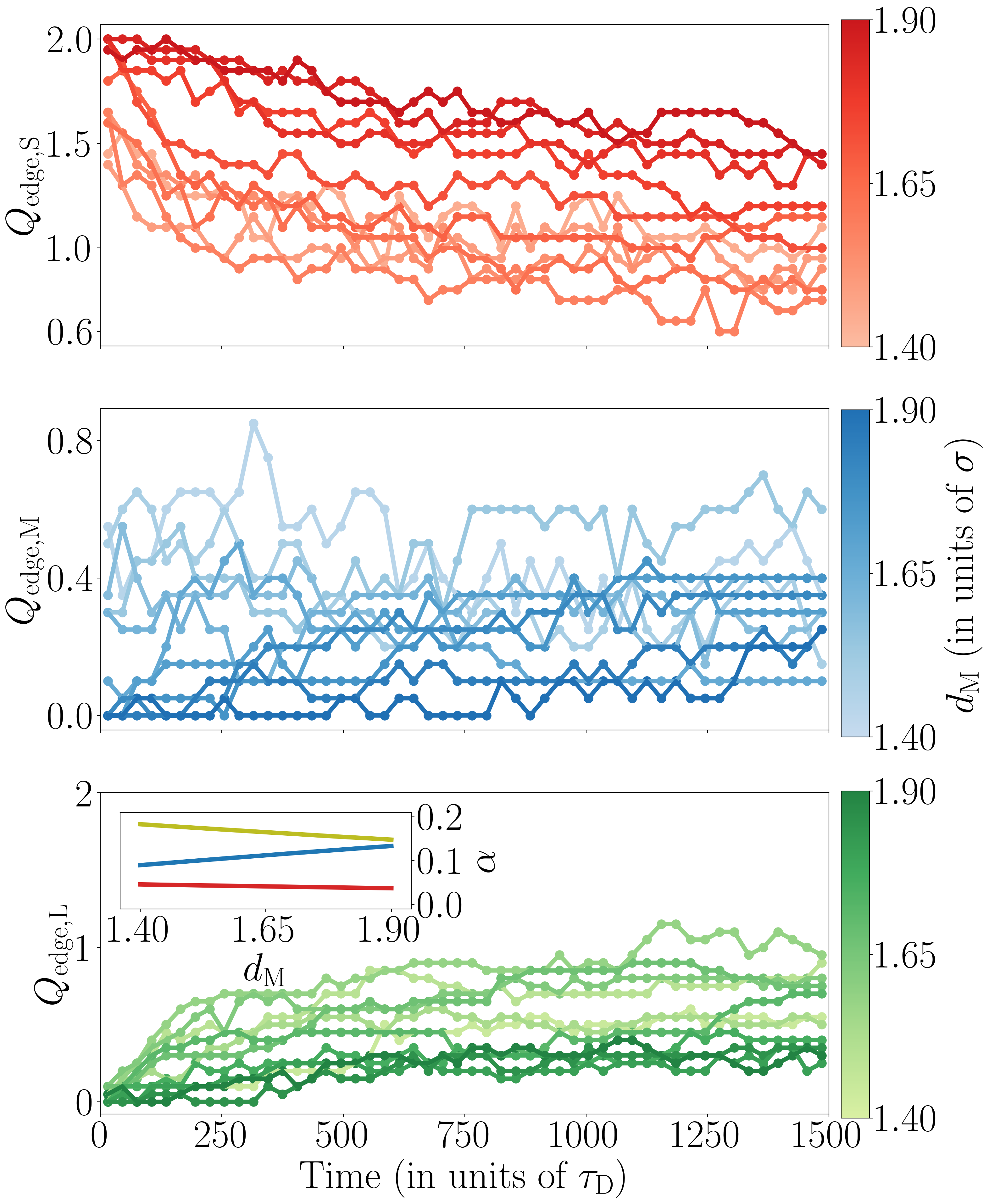}
    \caption{Number of small $Q_\mathrm{boundary,S}$, medium $Q_\mathrm{boundary,M}$ and large $Q_\mathrm{boundary,L}$ boundary lanes in dependence of the simulation time $t$ for the medium-sized particle diameter $d_\mathrm{M}\in[1.40,1.90]$.}
    \label{fig:res:NumberEdgeOverTS_ter}
\end{figure}

As depicted in Fig.~\ref{fig:res:funnelPic}, the funneling effect is based on a size difference between the involved particles, thus we varied the medium-sized particle diameter in the range $d_\mathrm{M}\in[1.40,1.90]$ to verify the influence onto the respective number of boundary lanes in ternary \textit{systems with identical particle number}. The number of boundary lanes over the simulation time is shown in Fig.~\ref{fig:res:NumberEdgeOverTS_ter} for each particle type and the driving force $F_\mathrm{S}=100$. At the beginning of the simulation, the boundary lanes are formed by exclusively the small and medium-sized particles, whereby significantly more boundary lanes consist of the smallest particles. For smaller medium-sized particle diameters $d_\mathrm{M}$, the amount of medium-sized and large particle boundary lanes $Q_\mathrm{boundary,M/L}$ increases. And the amount of small particle boundary lanes $Q_\mathrm{boundary,S}$ decreases. This dependency on the medium-sized particle diameter is consistent with the funneling effect explained in Fig.~\ref{fig:res:funnelPic}, since the width of the funnel area depicted in it depends on the diameter difference between the involved particle types. The larger the size difference, the larger the area in which a collision would result in exclusively a displacement of the smaller particle towards the wall. Consequently, more of the initially randomly distributed medium-sized particles are located in a funnel area, which results in thicker boundary lanes.\\
Just as observed for binary systems, the small particle boundary lanes are replaced by large particle boundary lanes with ongoing simulation time. The inset plot of Fig.~\ref{fig:res:NumberEdgeOverTS_ter} shows the area fraction of each particle type in dependence on the medium-sized particle diameter. Regardless of the diameter $d_\mathrm{M}$, the area fraction $\alpha_\mathrm{S}$ is the smallest. As observed in binary systems, a low area fraction results in thin boundary lanes of the respective particle type at early simulation times. Those thin boundary lanes are unstable over time and eventually vanish. Since the area fraction $\alpha_\mathrm{S}$ is comparable to the area fraction used in binary systems, the same explanation for the decrease of small particle boundary lanes with time is applicable for ternary systems.

\subsection{Density inhomogenities along the channel length}
\begin{figure*}
    \centering
    \includegraphics[width=.89\textwidth]{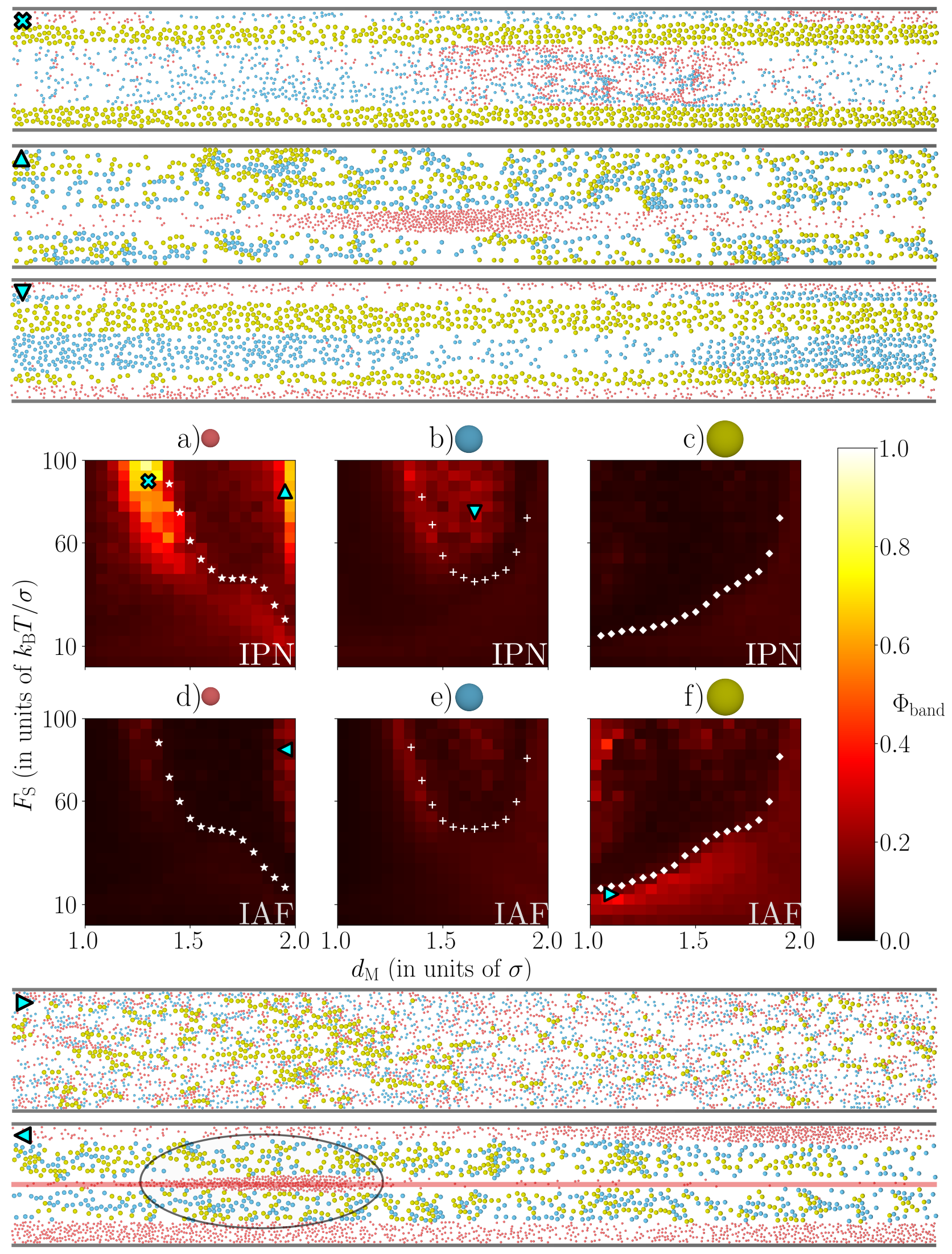}
    \caption{Band formation order parameter $\Phi_\mathrm{band}$ for the small, medium and large particles depicted as heat map. Panels a)-c) shows the parameter for IPN and d)-f) for IAF. The respective critical forces are plotted with white markers. The blue cross refers to a system with $a_\mathrm{M}=1.30$ and $F_\mathrm{S}=90$, the up triangle to $a_\mathrm{M}=1.95$ and $F_\mathrm{S}=85$, the down triangle to $a_\mathrm{M}=1.65$ and $F_\mathrm{S}=75$, the left triangle $a_\mathrm{M}=1.95$ and $F_\mathrm{S}=85$ and the right triangle to $a_\mathrm{M}=1.10$ and $F_\mathrm{S}=15$. For each of those markers, an exemplary snapshot of the channel is presented.}
    \label{fig:res:bandform}
\end{figure*}

Aside from the demixing into lanes, we observed a structure formation orthogonal to the driving force $F_\mathrm{S}$. The order parameter $\Phi_\mathrm{band}$ defined in Sec.~\ref{sec:observables} gives a numerical value representing the nonuniformity of the particle distribution in $x$-direction. Note that the phenomenon of band formation observed in Refs.~\cite{VaterMarc,Vissers2011_band} differs from the structures in this work, since we do not observe periodically repeating accumulations of particles. Therefore, we refer to a segregation of particles of one particle type orthogonal to the drive direction as a density inhomogeneity rather than a band. However, the band formation parameter is well suited to quantify these.\\
Fig.~\ref{fig:res:bandform} shows a heat map of the band formation parameter $\Phi_\mathrm{band}$ for ternary \textit{systems with identical particle number} (IPN) a)-c) and \textit{systems with identical area fraction} (IAF) d)-f) ranging from $\Phi_\mathrm{band}=0$ (dark) to $\Phi_\mathrm{band}=1.0$ (bright) for all three particle types. The white data points display the respective critical forces of lane formation (see Fig.~\ref{fig:res:terlanepics140}~e)) for IPN and Appendix C for IAF). We can distinguish between two different structures leading to density inhomogenities. \\
For the first one, the parameter $\Phi_\mathrm{band}$ is increased in systems, where the driving forces are just below the respective critical forces. This is particularly noticeable in Fig.~\ref{fig:res:bandform}~a) and f). The snapshot labeled with the blue cross refers to a ternary IPN with $a_\mathrm{M}=1.30$ and the driving force $F_\mathrm{S}=90$ (see Fig.~\ref{fig:res:bandform}~a)). At this driving force, the largest particles are segregated into lanes, while the small and medium-sized particles move in shared lanes. Inside of one of these shared lanes, the small particles and medium-sized particles move at different velocities. Hence, the transport of the faster, medium-sized particles is hindered by the slower, small particles. This could result in the accumulation in $x$-direction of the small particles. The same phenomenon is visible for the large particles in the snapshot labeled with the triangle pointing right (see Fig.~\ref{fig:res:bandform}~f)). Here, no particle type formed lanes, but the largest particles formed small chains and a slight increase of the driving force $F_\mathrm{S}$ induces lane formation of the large particles. The reason why this density inhomogeneity only exists for the small and large particles in Fig.~\ref{fig:res:bandform}~a) and f) is the method with which the number of particles in the channel is defined (IPN and IAF). The smaller particles display the behavior in Fig.~\ref{fig:res:bandform}~a) since they occupy the smallest area in a ternary IPN (see inset Fig.~\ref{fig:res:NumberEdgeOverTS_ter}). For example, in the snapshot labeled with the blue cross, the medium-sized particles are distributed uniformly through the whole channel length while at the same time a small increase of the driving force would result in the formation of small and medium-sized particle lanes. The band parameter is therefore not increased for the medium-sized particles, since they occupy considerably more channel area than the small particles. A similar explanation can be used for Fig.~\ref{fig:res:bandform}~f) and the snapshot labeled with the triangle pointing right. Here, the area fraction of each particle type is the same. Thus, the number of large particles is lower than the number of small or medium-sized particles. The low number of particles requires a higher driving force in order to connect the short chains to lanes which span the whole channel length. This results in a density inhomogeneity just below the respective critical forces.\\
In Fig.~\ref{fig:res:bandform}~a), b) and d), another phenomenon can be seen that results in the non-uniform distribution of particles. Here, density inhomogenities appear besides lane formation of the respective particle type. The snapshots labeled with the triangles pointing up and left show examples of this phenomenon for the small particles, and the snapshot labeled with the triangle pointing down for the medium-sized particles. In each example snapshot, the respective particles are non-uniformly distributed in $x$-direction, while moving inside their lane. For the small particles, this is most recognizable in the snapshot labeled with the triangle pointing left. The collection of small particles (see black circle) is asymmetrical. The left end is shaped more like wedge than the right end. This asymmetry can be explained by the different drift velocities in $x$-direction of the particles. Since the small particles are the slowest, the medium-sized and large particles pass from left to right. Now it becomes clear that the collection of small particles acts like a wedge on the left end, which creates a "slipstream". Inside this "slipstream", other small particles can catch up and join the collection. The medium-sized and large particles are diffusing into the empty space. This can be seen from the fact that the slipstream becomes thinner, the further it is from the wedge (see red shaded area in Fig.~\ref{fig:res:bandform} triangle pointing left).

\section{Conclusion \label{sec:conclusion}}
In this work, we investigated the lane and band formation of colloids modeled by Brownian dynamics simulations in binary and ternary systems. Hard walls confine the colloids inside a linear channel. The driving forces of the particle types differ, which was achieved by a difference in their size.\\
We found lane formation in ternary system. With increasing driving force, the particle types demix into lanes. A critical force indicating if and at which driving force lane formation is present in a particular system was defined. We found that the critical forces of all particle types depend on the diameter of the medium-sized particles $d_\mathrm{M}$. If two particle species of a system are similar in size, a high driving force is necessary to segregate them into lanes. In addition, we derived and confirmed a minimum of the medium-sized particle critical force at $d_\mathrm{M,min}=\sqrt[3]{(d_\mathrm{L}^3+d_\mathrm{S}^3)/2}$. With the values used in this work, the diameter results in $d_\mathrm{M,min}\approx 1.65$.\\
By increasing the simulation time frame to $1.5\cdot 10^3\tau_\mathrm{D}$, we expanded our understanding of the funneling effect in binary channels, which was introduced in our work Ref.~\cite{MarcKay}. Here, small particles can wedge between the wall and larger particles. This results in boundary lanes consisting of the smallest particles on short and medium time frames. We found that the stability of boundary lanes formed by the funneling effect depends on the width of the lanes themselves. Boundary lanes below a width of $2$ dissolve and vanish with time. This results in an increase of large particle boundary lanes. In addition, we found that the number of large particle boundary lanes depends on the small particle area fraction $\alpha_\mathrm{S}$. The higher the area fraction $\alpha_\mathrm{S}$ is, the fewer large particle boundary lanes exist.\\
In ternary systems, the funneling effect also results in mainly small particle boundary lanes, which are partly replaced by medium-sized or large particle boundary lanes over time. Furthermore, we found that the number of boundary lanes of each particle type depends on the medium-sized particle diameter $d_\mathrm{M}$. The bigger the medium-sized particles are, the more small particle boundary lanes and fewer medium-sized and large particle boundary lanes form.\\
Lastly, we studied the band formation parameter for ternary systems. We found two different phenomena which result in an increased band formation order parameter. First, at the driving force just below the critical force, the small and large particles form short chains. When the area fraction of the respective particle type is too low, the band formation parameter is increased. Second, at driving forces above the respective critical force, an increased band formation parameter occurs besides lane formation. This was presented for small and medium-sized particles. Here, the particles are non-uniformly distributed inside their lane, which is caused by mutual blocking of the particles.\\
In further works, it could be interesting to study other geometries (i.e. ring geometries) and check if the funneling effect persists there. Moreover, such a ring geometry might be a method to realize the periodic boundary conditions used in this work within an experimental setting. This approach could increase the experimentally viable observation time to study phenomena on long time scales in experiments. In addition, it could be interesting to study the size effect of the colloids decoupled from the driving force by using colloids with the same diameter but different driving forces. This could eliminate the small offset in Fig.~\ref{fig:res:terlanepics140}~e).

\section{ACKNOWLEDGMENTS~\label{sec:acknowledgments}}
This research was funded by the Deutsche Forschungsgemeinschaft DFG (Project No. NI259/16-1, ER341/13-1 and LE315/27-1). The authors gratefully acknowledge the Gauss Centre for Supercomputing e.V. (www.gauss-centre.eu) for funding this project by providing computing time through the John von Neumann Institute for Computing (NIC) on the GCS Supercomputer JUWELS at J\"ulich Supercomputing Centre (JSC).

\bibliography{lit}
\bibliographystyle{apsrev4-2}

\clearpage
\appendix
\section{Dependence on the channel length}
\begin{figure}[h]
    \centering
    \includegraphics[width=.4\textwidth]{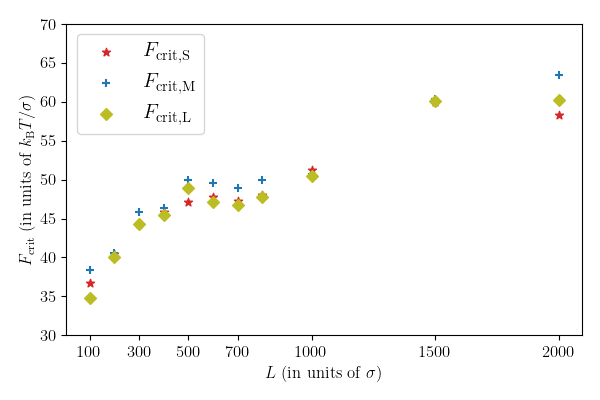}
    \caption{Critical forces $F_\mathrm{crit,S}$, $F_\mathrm{crit,M}$, and $F_\mathrm{crit,L}$ in dependence of the channel length $L$ for a medium-sized particle diameter $a_\mathrm{M}=1.65$. The critical force of each particle type increase for longer channels.}
    \label{fig:res:finitsize}
\end{figure}

To study the influence of the finite channel length $L$ on the critical force of each particle type, we fixed the medium-sized particle diameter $d_\mathrm{M}=1.65$ and varied the channel length $L$. Fig. \ref{fig:res:finitsize} shows these dependencies for each particle type. For short channels, $L<500$ we observe a rapid increase of the critical forces. This is a consequence of the periodic boundary conditions, where any demixing into lanes at the end of the channel is passed to the beginning. Thus, in short channels, the lane formation process dominates over diffusion even for low driving forces. In the range of, $400<L<1000$ the critical force of each particle type is roughly constant at below $F_\mathrm{crit}=50$. For long channels, $L>900$ the critical force increases again, except with a smaller slope than for short channels. To obtain a reasonable computation time, we fixed the channel length for all other simulations to $L=400$, since the rapid increase of the critical forces begins to flatten at this length.

\section{Concerning the stationary state}
\begin{figure}[h]
    \centering
    \includegraphics[width=.5\textwidth]{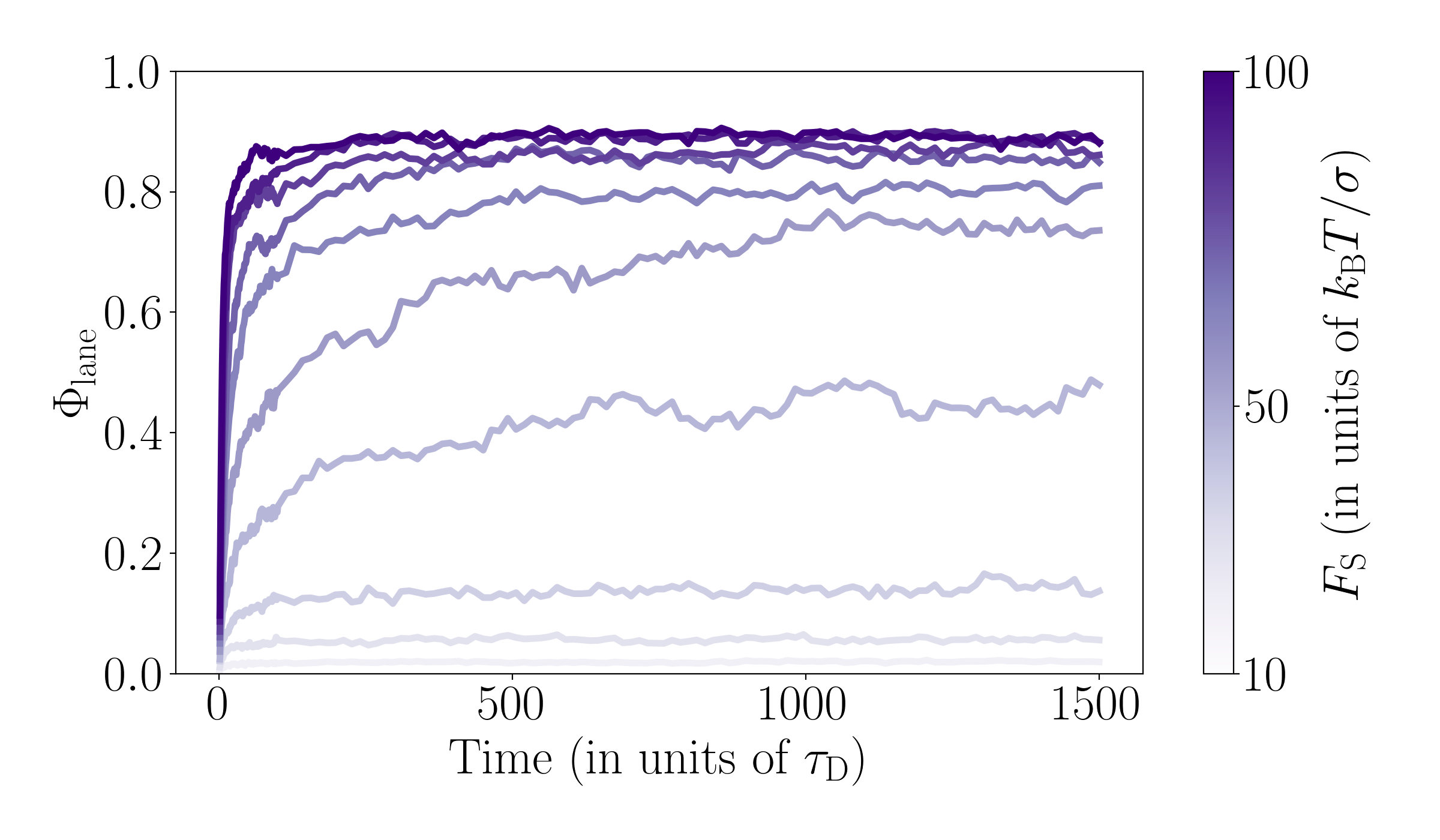}
    \caption{Lane formation order parameter $\Phi_\mathrm{lane}$ in dependence of the simulation duration $t$. The process of lane formation is very rapid, and the order parameter reached a plateau after $t=500$ for every driving force.}
    \label{fig:res:statState}
\end{figure}

In Sec. IV. A. and B. the order parameter $\Phi_\mathrm{lane}$ is used to calculate the critical forces. To ensure that a stationary state is reached in the ternary systems, we checked the time evolution of the order parameter $\Phi_\mathrm{lane}$ averaged over all particle types until $t=1500$. This is presented in Fig.\ref{fig:res:statState} for a ternary system with a medium-sized particle diameter $d_\mathrm{M}=1.65$ and a channel length $L=400$. The number of particles of each type is equal. The different driving forces are color coded and range from $F_\mathrm{S}=10$ (light blue) to $F_\mathrm{S}=100$ (dark blue) in steps of $\Delta F_\mathrm{S}=10$. Lane formation occurs rapidly, with the $\Phi_\mathrm{lane}$ parameter plateauing the latest at approximately $t=500$ for all driving forces. The higher the driving force, the early the plateauing of the lane formation parameter.

\section{Lane formation of ternary systems with identical area fraction}
\begin{figure}
    \centering
    \includegraphics[width=.5\textwidth]{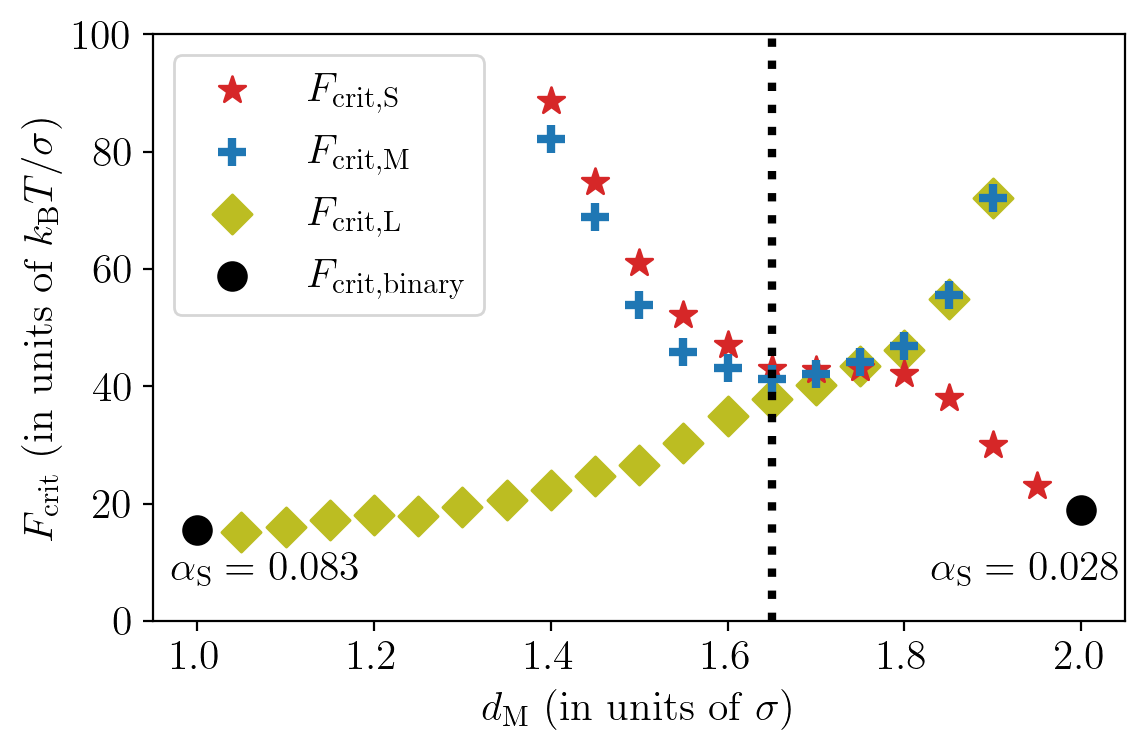}
    \caption{Dependence of the critical force of each particle type on the medium-sized particle diameter $d_\mathrm{M}$ for an IPN. The black circles correspond to binary systems. The dotted black line marks $d_\mathrm{M}=1.65$}
    \label{fig:res:fcrit}
\end{figure}

\begin{figure}
    \centering
    \includegraphics[width=.5\textwidth]{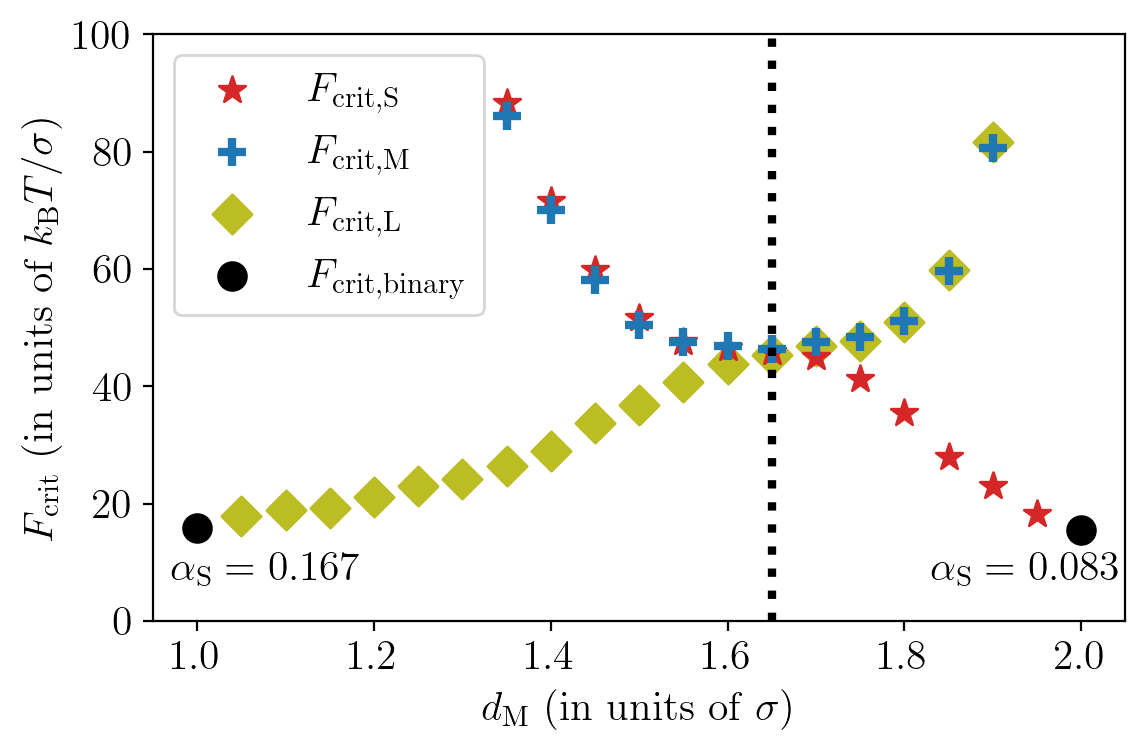}
    \caption{Dependence of the critical force of each particle type on the medium-sized particle diameter $d_\mathrm{M}$ for an IAF. The black circles correspond to binary systems. The dotted black line marks $d_\mathrm{M}=1.65$}
    \label{fig:res:fcrit2}
\end{figure}

Besides the IPN (\textit{systems with identical particle number}), we studied IAF (\textit{systems with identical area fraction}) as well. Regarding lane formation, we did not find many differences between those systems types. This is clear by comparing Fig. \ref{fig:res:fcrit} (same as Fig. 4 b) in the manuscript) with Fig. \ref{fig:res:fcrit2}, which shows the critical force of each particle type $F_\mathrm{crit}$ vs. the medium-sized particle diameter $d_\mathrm{M}$. The dashed black line depicts the derived value $d_\mathrm{M,min}=1.65$. The black dots represent binary systems. However, two differences are noticeable. First, the critical force of the largest particle show a saddle point at around $d_\mathrm{M,min}=1.65$ in the IAF, which is not visible in the IPN. Secondly, the derived value $d_\mathrm{M,min}=1.65$ provides a more precise description of the critical force minima for the medium-sized particles in the IAF system.

\section{Comparative analysis of observation times in experimental setups and simulations}
In this section, we present an exemplary calculation of the attainable observation time within a typical experimental setup. The experiment involves a channel of length $L_\mathrm{exp}$ filled with colloids of diameter $d_\mathrm{exp}$ suspended in a solvent. The colloids are introduced at one end of the channel and, due to the inclination of the channel, gravity propels these colloids along its length. After traversing the entire length $L_\mathrm{exp}$, the colloids leave the channel. Therefore, in an experimental setting, the observation time for lane formation is limited to the time a colloid needs to move through the length of the channel. The time a particle needs to traverse the channel can be roughly estimated with the help of the mean drift velocity $\langle v\rangle_\mathrm{D}=DF_\mathrm{exp}/k_\mathrm{B}T$. Here, $k_\mathrm{B}T$ is the thermal energy, $D=k_\mathrm{B}T/3\pi\eta d_\mathrm{exp}$ the diffusion constant with the viscosity of the solvent $\eta$, and the driving force due to gravity $F_\mathrm{exp}$. In order to obtain a value for the driving force, we can convert values given in this work for the critical force to SI-units with $F_\mathrm{exp}=F_\mathrm{crit}k_\mathrm{B}T/d_\mathrm{exp}$. In total, the time a colloid needs to move through the channel length $L$ results to
\begin{align*}
    T_\mathrm{exp}=\frac{L_\mathrm{exp}}{\langle v\rangle_D}=\frac{3\pi L_\mathrm{exp}d_\mathrm{exp}^2\eta}{F_\mathrm{crit}k_\mathrm{B}T} \quad .
\end{align*}
In accordance to prior experiments, we selected $L=\SI{2}{\milli\metre}$ [29] and $d_\mathrm{exp}=\SI{4}{\micro\metre}$ [31]. For simplicity, we chose water as the solvent at room temperature $T=\SI{293.15}{\kelvin}$, which determines the viscosity $\eta=\SI{1}{\milli\pascal\second}$ [30]. Given these parameters and a critical force of $F_\mathrm{crit}=40$ (which approximately corresponds to the crossing point for $d_\mathrm{M} = 1.65$ in Fig.~4b) in the main manuscript), the colloid requires $T_\mathrm{exp}=\SI{1863}{\second}\approx\SI{0.5}{\hour}$ to traverse through the channel. This confines the experimental observation of lane formation phenomena roughly to this time scale.\\
In computer simulations, we can avoid this temporal constraint by the use of periodic boundary conditions. Here, the longest simulation time is $T_\mathrm{sim}=1500\tau_\mathrm{D}=1500( d_\mathrm{exp}^2/D)\approx\SI{63}{\hour}$. By comparing these time scales, it is evident that $T_\mathrm{exp}\ll T_\mathrm{sim}$. We suppose that the results of this work, which only appear on very long time scales, can not be validated using such an experimental setup.

\end{document}